\documentclass[a4paper]{spie}  

\usepackage{amsmath,amsfonts,amssymb}
\usepackage{graphicx}
\usepackage[colorlinks=true, allcolors=blue]{hyperref}

\def \mt {}

\title{The e-ASTROGAM gamma-ray space mission}

\author[a]{V. Tatischeff}
\author[b,c,d]{M. Tavani}
\author[e]{P. von Ballmoos}
\author[f]{L. Hanlon}
\author[g]{U. Oberlack}
\author[h]{A. Aboudan}
\author[b]{A. Argan}
\author[i]{D. Bernard}
\author[g]{A. Brogna}
\author[h]{A. Bulgarelli}
\author[j]{A. Bykov}
\author[h]{R. Campana}
\author[k]{P. Caraveo}
\author[l]{M. Cardillo}
\author[m]{P. Coppi}
\author[n]{A. De Angelis}
\author[o]{R. Diehl}
\author[b]{I. Donnarumma}
\author[h]{V. Fioretti}
\author[k]{A. Giuliani}
\author[p]{I. Grenier}
\author[q]{J. E. Grove}
\author[a]{C. Hamadache}
\author[r]{D. Hartmann}
\author[s]{M. Hernanz}
\author[s]{J. Isern}
\author[o]{G. Kanbach}
\author[a]{J. Kiener}
\author[e]{J. Kn\"odlseder}
\author[h]{C. Labanti}
\author[t]{P. Laurent}
\author[u]{O. Limousin}
\author[v]{F. Longo}
\author[h]{M. Marisaldi}
\author[f]{S. McBreen}
\author[w]{J. E. McEnery}
\author[k]{S. Mereghetti}
\author[u]{F. Mirabel}
\author[x]{A. Morselli}
\author[y]{K. Nakazawa}
\author[a]{J. Peyr\'e}
\author[b]{G. Piano}
\author[z,aa]{C. Pittori}
\author[b]{S. Sabatini}
\author[ab,ac]{L. Stawarz}
\author[w]{D. J. Thompson}
\author[f]{A. Ulyanov}
\author[ad]{R. Walter}
\author[ae]{X. Wu}
\author[af]{A. Zdziarski}
\author[ag]{A. Zoglauer}

\affil[a]{CSNSM, IN2P3-CNRS and Univ Paris-Sud, F-91405 Orsay Cedex, France}
\affil[b]{INAF-IAPS, via del Fosso del Cavaliere 100, I-00133 Roma, Italy}
\affil[c]{Dip. di Fisica, Univ. Roma Tor Vergata, via della Ricerca Scientifica 1, I-00133 Roma, Italy}
\affil[d]{Gran Sasso Science Institute, viale Francesco Crispi 7, I-67100 L'Aquila, Italy}
\affil[e]{IRAP, 9, av. du Colonel-Roche, Toulouse 31028, France}
\affil[f]{School of Physics, University College Dublin, Belfield, Dublin 4, Ireland}
\affil[g]{PRISMA Detector Laboratory, Johannes Gutenberg Univ. Mainz, Germany}
\affil[h]{INAF-IASF-Bologna, Via Gobetti 101, I-40129 Bologna, Italy}
\affil[i]{LLR, Ecole Polytechnique, CNRS/IN2P3, 91128 Palaiseau, France}
\affil[j]{Ioffe Institute, Saint-Petersburg, 194021, Russian Federation}
\affil[k]{INAF-IASF Milano, via E. Bassini 15, I-20133 Milano, Italy}
\affil[l]{INAF Osservatorio Astronomico di Arcetri, Largo Enrico Fermi, 5, I-50125 Firenze, Italy}
\affil[m]{Department of Astronomy, Yale University, P.O. Box 208101, New Haven, 06520-8101 USA}
\affil[n]{INFN Padova, via Marzolo 8, 35141, Padova, Italy}
\affil[o]{Max Planck Institut f\"ur extraterrestrische Physik, D-85748 Garching, Germany}
\affil[p]{AIM Paris-Saclay, CEA/IRFU, CNRS, Univ Paris Diderot, 91191 Gif-sur-Yvette, France}
\affil[q]{Space Science Division, Naval Research Laboratory, Washington, DC 20375-5352, USA}
\affil[r]{Department of Physics and Astronomy, Clemson University, Clemson, SC 29634, USA}
\affil[s]{ICE-CSIC/IEEC, Campus UAB, 08193 Bellaterra, Barcelona, Spain}
\affil[t]{APC, Univ Paris Diderot, CNRS/IN2P3, CEA/Irfu, Observatoire de Paris, 10 rue Alice Domont et L\'eonie Duquet, F-75205 Paris Cedex 13, France}
\affil[u]{CEA Saclay, DSM/Irfu/Service d'Astrophysique, 91191 Gif-sur-Yvette Cedex, France}
\affil[v]{Dip. di Fisica, Univ. di Trieste and INFN, Via Valerio 2, I-34127 Trieste, Italy}
\affil[w]{NASA Goddard Space Flight Center, Greenbelt, MD 20771, USA}
\affil[x]{INFN Roma Tor Vergata, via della Ricerca Scientifica 1, I-00133 Roma, Italy}
\affil[y]{Department of Physics, The University of Tokyo, 7-3-1 Hongo, Bunkyo-ku, Tokyo, Japan}
\affil[z]{ASI Science Data Center (ASDC), Via del Politecnico, I-00133 Roma, Italy}
\affil[aa]{INAF-OAR, via Frascati 33, I-00078 Monte Porzio Catone (Roma), Italy}
\affil[ab]{ISAS, JAXA, 3-1-1 Yoshinodai, Chuo-ku, Sagamihara, Kanagawa 252-5210, Japan}
\affil[ac]{Astronomical Observatory, Jagiellonian University, ulica Orla 171, 30-244 Krak\`ow, Poland}
\affil[ad]{ISDC, Geneva Observatory, University of Geneva, Switzerland}
\affil[ae]{D\'epartement de Physique Nucl\'eaire et Corpusculaire, University of Geneva, Switzerland}
\affil[af]{Centrum Astronomiczne im. M. Kopernika, Bartycka 18, PL-00-716 Warszawa, Poland}
\affil[ag]{Space Sciences Laboratory, University of California, Berkeley, CA 94720, USA}

\authorinfo{On behalf of the e-ASTROGAM Collaboration (see \url{http://astrogam.iaps.inaf.it/}). \\
Corresponding author e-mail address: Vincent.Tatischeff@csnsm.in2p3.fr
}

\begin{document}
\maketitle

\begin{abstract}
e-ASTROGAM is a  gamma-ray {\mt space mission} 
to be proposed as the M5 Medium-size mission of the European Space
Agency. It is dedicated to the observation of the Universe with
unprecedented sensitivity in the energy range 0.2 -- 100 MeV,
extending up to GeV energies, together with a groundbreaking
polarization capability. {\mt It is designed to substantially
improve the COMPTEL and Fermi sensitivities in the MeV-GeV energy
range and to open new windows of opportunity for astrophysical and
fundamental physics space research.} e-ASTROGAM will operate as an
open astronomical observatory, with a core science focused on (1)
the activity from extreme particle accelerators, including
gamma-ray bursts and active galactic nuclei and the link of jet
astrophysics to {\mt the new astronomy of gravitational waves,
neutrinos, ultra-high energy cosmic rays}, (2) the high-energy
mysteries of the Galactic center and inner Galaxy, including the
activity of the supermassive black hole, the {\it Fermi} Bubbles,
the origin of the Galactic positrons, and the search for dark
matter signatures in a new energy window; (3) nucleosynthesis and
chemical evolution, including the life cycle of elements produced
by supernovae in the Milky Way and the Local Group of galaxies.
e-ASTROGAM will be ideal for the study of high-energy sources in
general, including pulsars and pulsar wind nebulae, accreting
neutron stars and black holes, novae, supernova remnants, and
magnetars. And it will also provide important contributions to
solar and terrestrial physics. The e-ASTROGAM
telescope {\mt is}  
optimized for the simultaneous
detection of Compton and pair-producing gamma-ray events over a
large spectral band. It is based on a very high technology
readiness level for all subsystems and includes many innovative
features for the detectors and associated electronics.
\end{abstract}


\section{INTRODUCTION}
\label{sec:intro}  

During the last decade, gamma-ray astronomy has experienced a
period of impressive {\mt scientific advances and successes.} 
In the high-energy range {\mt above 100 MeV, the {\it AGILE} mission first 
and then the {\it Fermi} mission led to important discoveries. In
particular,} 
the Large Area Telescope (LAT) of the {\it Fermi} satellite has
established an inventory of over 3000 steady sources of various
kinds (blazars, pulsars, supernova remnants, high-mass binaries,
{\mt gamma-ray bursts}, etc.) showing a variety of gamma-ray emission
processes\cite{3FGL} . Similarly, in the hard X-ray/low-energy
gamma-ray band, the latest catalog of sources detected with the
Burst Alert Telescope of the {\it Swift} mission contains 1171
sources in the range 14--195~keV\cite{bau13}. But at intermediate
photon energies, between 0.2 and 100 MeV, only a few tens of
steady sources have been detected so far, {\mt mostly by the
COMPTEL instrument on board the {\it Compton Gamma-Ray Observatory}} (CGRO; see Ref.\cite{sch00}), such
that this peculiar field of astronomy has remained largely
unexplored.

However, many of the most-spectacular objects in the Universe have
their peak emissivity at photon energies between 0.2 and 100 MeV
(e.g. gamma-ray bursts, blazars, pulsars, etc.), so it is in this
energy band that essential physical properties of these objects
can be studied most directly. This energy range is also known to
feature a characteristic spectral turn-over associated to hadronic
emission from pion decay, which makes it paramount for the study
of the radiating, nonthermal particles {\mt and for distinguishing
leptonic from hadronic processes}. Moreover, this energy domain
covers the crucial range of nuclear gamma-ray lines produced by
radioactive decay, nuclear collision, positron annihilation, or
neutron capture, which makes it as special for high-energy
astronomy as optical spectroscopy is for phenomena related to atomic
physics.

The e-ASTROGAM mission concept {\mt originates from the ASTROGAM
mission proposed in early 2015 for the ESA M4 call (see
Ref.\cite{tav16})}. {\mt The e-ASTROGAM gamma-ray instrument
inherits from predecessors such as {\it AGILE}\cite{tav09} and
{\it Fermi}\cite{atw09}, as well as from the MEGA
prototype\cite{kan05}, but it takes full advantages of recent
progresses in silicon detectors and readout microelectronics to
achieve measurement of the energy and 3D position of each
interaction within the detectors with an excellent spectral and
spatial resolution.} {\mt The main innovative feature of the
e-ASTROGAM mission is the capability of joint detection in the
Compton (0.2 -- 30 MeV) and pair ($>10$~MeV) energy ranges in a
single integrated instrument.}
 The mission aims {\mt at improving} 
 the sensitivity in the medium-energy
gamma-ray domain by one to two orders of magnitude compared to
previous missions. It is also intended to provide a groundbreaking
capability for measuring gamma-ray polarization, thus giving
access to a new observable that can provide valuable information
on the geometry and emission processes of various high-energy
sources. The proposed mission has the potential to answer key
questions in astrophysics through a dedicated core science
program:
\begin{enumerate}

\item {\bf Jet astrophysics in the era of time-domain astronomy {\mt and a
unique instrument for the new
astronomy of gravitational wave sources and other energetic transients} } 
{\it What is the mechanism responsible for the launch of
ultra-relativistic jets in gamma-ray bursts (GRBs)? What is the
composition of these jets and how is the prompt GRB emission 
produced? Can short-duration GRBs be {\mt unequivocally}
associated to gravitational wave signals? How does the accretion
disk/jet transition occur around super-massive black holes in
active galactic nuclei (AGN)? Are blazars of the BL Lacertae type
sources of ultra-high energy cosmic rays (UHECRs) and high-energy
neutrinos?} With its wide field of view (\S~3.2), e-ASTROGAM will
give access to a variety of high-energy, transient phenomena
evolving on time scales ranging from milliseconds to years. With
its unprecedented sensitivity over a large energy range (\S~3.4),
combined with an exceptional capacity for polarimetry (\S~3.6),
e-ASTROGAM will provide unambiguous answers to these questions for
the first time.

\item {\bf The high-energy mysteries of the Galactic Center}. {\mt
The Galactic Center (GC) region hosts a super-massive black hole
and a variety of compact objects of mysterious nature. Their
emission mechanisms and feedback on the Galactic environment are
still poorly understood.} {\it What is the origin of the gigantic,
gamma-ray emitting bubbles (the Fermi Bubbles) that extend about
25,000 light-years north and south of the galactic center? Is
there a link between the observed annihilation of a tremendous
amount of positrons in the Galaxy's bulge, the Fermi Bubbles, and
the activity of the central supermassive black hole? What is
causing the GeV excess emission detected from the Galactic center
region?} With an angular resolution significantly improved over
previous missions (\S~3.3), e-ASTROGAM will enable spectro-imaging
of the Fermi Bubbles at low and medium latitudes in an energy band
(0.2~MeV~--~3~GeV) crucial to distinguish between models. e-ASTROGAM
will also make great strides in positron astrophysics by producing
maps of annihilation radiation far superior to those of {\it
INTEGRAL}/SPI, thanks to its much better sensitivity at 511~keV
(by a factor of $13$ for the narrow line component; see \S~3.5)
and spatial resolution. {\mt Furthermore, the e-ASTROGAM
unprecedented sensitivity and angular resolution in the MeV~--~GeV
range would uniquely contribute to dark matter studies and
resolution of current problems regarding the role of cosmic rays
in shaping star formation and Galactic properties.}

\item {\bf Supernovae, nucleosynthesis, and Galactic chemical evolution}.
{\it How do thermonuclear and core-collapse supernovae (SNe) explode?
How are cosmic isotopes created in stars and distributed in the interstellar medium?}
e-ASTROGAM will achieve a remarkable improvement in line sensitivity over previous missions
(\S~3.5), thus enabling a decisive progress in the field of astronomy with radioactivities.
The mission should allow us to finally understand the progenitor system(s) and explosion
mechanism(s) of Type Ia SNe, the dynamics of core collapse in massive star explosions,
as well as the history of recent SNe in the Milky Way.

\end{enumerate}
In these proceedings, we first provide a description of the e-ASTROGAM gamma-ray telescope and then discuss the  instrument performance in the light of these scientific objectives.

\section{THE e-ASTROGAM TELESCOPE}
\label{sec:instrument}

\begin{figure}
\centering
\includegraphics[width=0.55\textwidth]{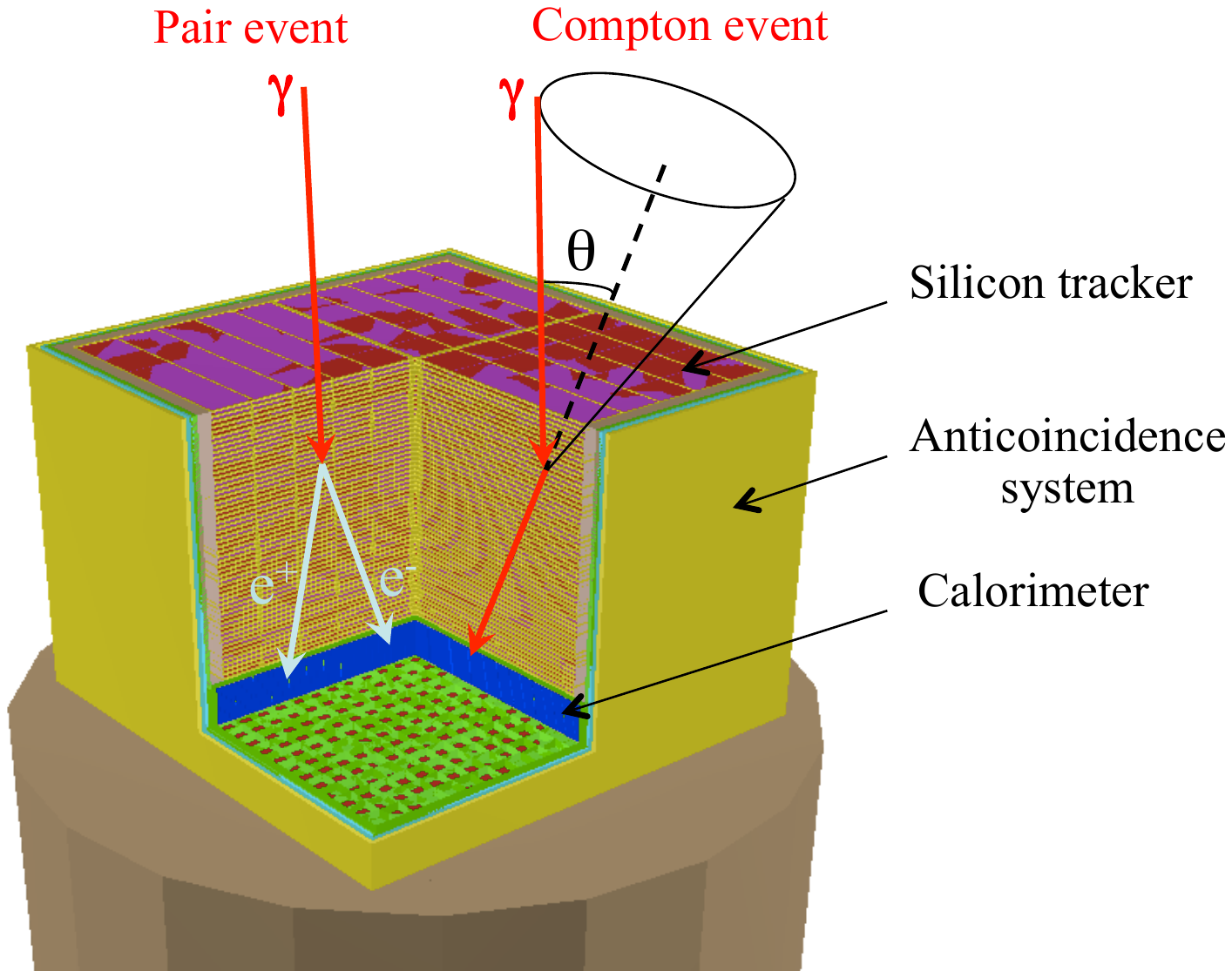}
\caption{Geant4/MEGAlib mass model of the e-ASTROGAM gamma-ray telescope on the satellite
platform, together with a representation of a pair-creation event and of a Compton event. The Anticoincidence detector of the top face is not shown in this figure.}
\label{fig:massmodel}
\end{figure}

The e-ASTROGAM telescope is made up of three detection system: a
silicon {\bf Tracker} in which the cosmic gamma rays undergo a Compton scattering
or a pair conversion, a {\bf Calorimeter} to absorb and measure the energy of the
secondary particles, and an {\bf anticoincidence} (AC) system to veto the prompt-reaction
background induced by charged particles (see Figure~\ref{fig:massmodel}). The telescope has
a size of 110$\times$110$\times$80 cm$^3$ and a mass of 820~kg (not including maturity margins).

\subsection{Silicon Tracker}

The Tracker is the heart of the e-ASTROGAM payload. It is based on the silicon strip detector technology widely employed in medical imaging and particle physics experiments (e.g. ATLAS and CMS at LHC), and already applied to the detection of gamma rays in space with the {\it AGILE}\cite{bul07} and {\it Fermi}\cite{atw07} missions. The e-ASTROGAM Tracker needs double-sided strip detectors (DSSDs), however, to work also as a Compton telescope.

The Si Tracker comprises 5600 DSSDs arranged in 56 layers (100 DSSDs per layer). It is divided in four units of 5$\times$5 DSSDs, the detectors being wire bonded strip to strip to form 2-D ladders. The interlayer distance is 10~mm. Each DSSD has a geometric area of 9.5$\times$9.5 cm$^2$, a thickness of 500~$\mu$m, and a strip pitch of 240~$\mu$m. The total detection area amounts to 9025 cm$^2$ and the total Si thickness to 2.8~cm, which corresponds to 0.3 radiation length on axis, and a probability of a Compton interaction at 1~MeV of 40\%. Such a stacking of relatively thin detectors enables an efficient tracking of the electrons and positrons produced by pair conversion, and of the recoil electrons produced by Compton scattering. The DSSD signals are readout by 860,160 independent, low-power electronics channels with self-triggering capability.

The essential characteristics of the e-ASTROGAM Tracker are (i) its charge readout with an excellent spectral resolution (about 6 keV FWHM noise level in the baseline configuration of the front-end electronics), allowing to accurately measure the low-energy deposits produced by Compton events, and (ii) its light mechanical structure minimizing the amount of passive material within the detection volume to enable the tracking of low-energy Compton electrons and electron-positron pairs, and improve the point spread function in both the Compton and pair domains by reducing the effect of multiple Coulomb scattering.

\subsection{Calorimeter}

The e-ASTROGAM Calorimeter is a pixelated detector made of a high-$Z$ scintillation material -- Thallium activated Cesium Iodide -- for an efficient absorption of Compton scattered gamma-rays and electron-positron pairs. It consists of an array of 33,856 parallelepiped bars of CsI(Tl) of 8~cm length and 5$\times$5~mm$^2$ cross section, read out by silicon drift detectors (SDDs) at both ends, arranged in an array of 529 ($=23 \times 23$) elementary modules comprising each 64 crystals. The Calorimeter thickness -- 8 cm of CsI(Tl) -- makes it a 4.3 radiation-length detector having an absorption probability of a 1-MeV photon on axis of 88\%.

The Calorimeter detection principle and architecture are based on the heritage of the space instruments {\it INTEGRAL}/PICsIT, {\it AGILE}/MCAL and {\it Fermi}/LAT, as well as on the particle physics experiment LHC/ALICE at CERN. However, the e-ASTROGAM calorimeter features two major improvements with respect to previous instruments: (i) The energy resolution is optimized to a FWHM of 4.5\% at 662 keV (scaling with the inverse of the square root of the energy) by the use of low-noise SDDs for the readout of the scintillation signals, combined with an appropriate ultra low-noise FEE, and (ii) the spatial resolution is improved by measuring the depth of interaction in the detector from a suitable weighting function of the recorded scintillation signals at both ends. Accurately measuring the 3D position and deposited energy of each interaction is essential for a proper reconstruction of the Compton events.

The simultaneous data set provided by the Silicon Tracker, the Calorimeter and the Anticoincidence system constitutes the basis for the gamma-ray detection. However, the Calorimeter will also have the capability to trigger the gamma-ray event processing independently of the Tracker, in order to search for fast transient events such as GRBs and terrestrial gamma-ray flashes.

\subsection{Anticoincidence system}

The third main detector of the e-ASTROGAM payload consists of an Anticoincidence (AC) system made of segmented panels of plastic scintillators covering the top and four lateral sides of the instrument, requiring a total active area of about 4.7~m$^2$. The AC detector is segmented in 33 plastic tiles (6 tiles per lateral side and 9 tiles for the top). All scintillator tiles are coupled to silicon photomultipliers (SiPM) by optical fibers. The architecture of the AC detector is fully derived from the successful design of the {\it AGILE} \cite{per06} and {\it Fermi}/LAT \cite{moi07} AC systems. In particular, their segmentation has proven successful at limiting the ``backsplash'' self-veto, therefore the dead time of the instrument. The AC particle background rejection is designed to achieve a relativistic charged particle detection inefficiency lower than 10$^{-4}$, a standard value already realized in current space experiments.

\section{TELESCOPE PERFORMANCE ASSESSMENT}
\label{sec:performance}

The numerical mass model of e-ASTROGAM used to simulate the performance of the instrument is shown in Figure~\ref{fig:massmodel}. The simulations were performed with the software tools MEGAlib and BoGEMMS. The MEGAlib package\cite{zog06} was originally developed for analysis of simulation and calibration data related to the Compton scattering and pair-creation telescope MEGA\cite{kan05}. It has then been successfully applied to a wide variety of hard X-ray and gamma-ray telescopes on ground and in space, such as COMPTEL, NCT, and {\it NuSTAR}. BoGEMMS (Bologna Geant4 Multi-Mission Simulator) is a software for simulation of payload of X- and gamma-ray missions, which has been developed at the INAF/IASF Bologna\cite{bul12}. It has already been applied to several hard X-ray/gamma-ray instruments and mission projects, including Simbol-X, NHXM, Gamma-Light, {\it AGILE}, and GAMMA-400. Both software packages exploit the Geant4 toolkit to model the geometrical and physical parameters of the detectors and simulate the interactions of photons and particles in the instrument.

\subsection{Background model}

For best environmental conditions, ASTROGAM should be launched into a quasi-equatorial (inclination $i < 2.5^\circ$)  low-Earth orbit (LEO) at a typical altitude of 550~km. The background environment in such an orbit is now well-known, thanks to the Beppo-SAX mission, which measured the radiation environment on a low-inclination ($i \sim 4^\circ$), 500 -- 600 km altitude orbit almost uninterruptedly during 1996 -- 2002 \cite{cam14} and the on-going {\it AGILE} mission, which scans the gamma-ray sky since 2007 from a quasi-equatorial ($i \sim 2.5^\circ$) LEO at an average altitude of 535~km \cite{tav09}. The dominant sources of background for the ASTROGAM telescope in the MeV domain are the cosmic diffuse gamma-ray background, the atmospheric gamma-ray emission, the reactions induced by albedo neutrons, and the background produced by the radioactivity of the satellite materials activated by fast protons and alpha particles. All these components were carefully modeled using the MEGAlib environment tools. In the pair domain above 10 MeV, the background is mainly induced by fast particles (mainly leptons) impinging the spacecraft, as well as by the cosmic diffuse radiation and the atmospheric gamma-ray emission.

\subsection{Field of view}

The e-ASTROGAM field of view  was evaluated from detailed simulations of the incident angle dependence of the sensitivity. In the Compton domain, the sensitivity remains high within $40^\circ$ to $50^\circ$ off-axis angle and then degrades for larger incident angles. For example, the field of view at 1~MeV amounts to 46$^\circ$ half width at half maximum (HWHM), with a fraction-of-sky coverage in zenith pointing mode of 23\%, corresponding to $\Omega = 2.9$~sr.

In the pair-production domain, the field-of-view assessment is also based on in-flight data from the {\it AGILE} and {\it Fermi}-LAT gamma-ray imager detectors. With the e-ASTROGAM characteristics (size, Si plane spacing, overall geometry), the field of view is found to be $> 2.5$~sr above 10~MeV.  

\subsection{Angular and spectral resolution}

\begin{figure}
\centering
\includegraphics[width=0.95\textwidth]{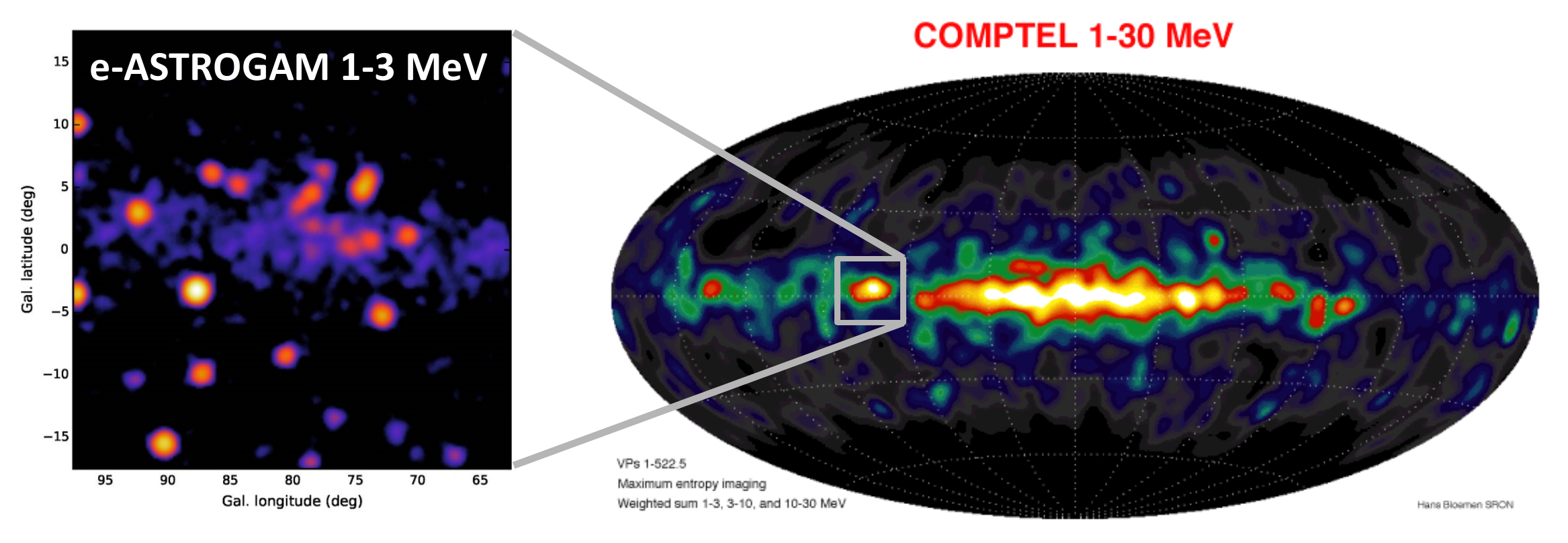}
\caption{{\it Right panel} -- CGRO/COMPTEL map of the Milky Way in the 1 to 30 MeV energy band (Credit: COMPTEL Collaboration). {\it Left panel} -- Simulated e-ASTROGAM map of the Cygnus region in the 1 -- 3 MeV band, from an extrapolation of the 3FGL source spectra to low energies (see text).}
\label{fig:psf}
\end{figure}

e-ASTROGAM will achieve an unprecedented angular resolution both in the MeV domain and above a few hundreds of MeV, i.e. improving the angular resolution of the {\it CGRO}/COMPTEL telescope and that of the {\it Fermi}/LAT instrument by a factor of $\sim$4 at 1 and 400~MeV, respectively.

In the pair production domain, the PSF improvement over {\it Fermi}/LAT is due to (i) the absence of heavy converters in the Tracker, (ii) the light mechanical structure of this detector minimizing the amount of passive material within the detection volume and thus enabling a better tracking of the secondary electrons and positrons, and (iii) the analog readout of the DSSD signals allowing a fine spatial resolution of about 40~$\mu$m ($\sim$1/6 of the microstrip pitch). In the Compton domain, thanks to the fine spatial and spectral resolutions of both the Tracker and the Calorimeter, the e-ASTROGAM angular resolution will be close to the physical limit induced by the Doppler broadening due to the velocity of the target atomic electrons.

Figure~\ref{fig:psf} shows an example of the e-ASTROGAM imaging capability in the MeV domain compared to COMPTEL. The e-ASTROGAM synthetic map of the Cygnus region was produced from the third {\it Fermi} LAT (3FGL) catalog of sources detected at photon energies $E_\gamma > 100$~MeV\cite{3FGL}, assuming a simple extrapolation of the measured power-law spectra to lower energies. It is clear from this example that e-ASTROGAM will substantially overcome (or eliminate in some cases) the confusion issue that severely affected the previous and current generations of gamma-ray telescopes. The e-ASTROGAM imaging potential will be particularly relevant to study the various high-energy phenomena occuring in the Galactic center region.

e-ASTROGAM will also significantly improve the energy resolution with respect to COMPTEL, e.g. by a factor of $\sim$3.2 at 1 MeV, where it will reach a 1$\sigma$ resolution of $\Delta E/E=1.3$\%. In the pair production domain above 30~MeV, the simulated spectral resolution is within 20--30\%.

\subsection{Continuum sensitivity}

\begin{figure}
\centering
\includegraphics[width=0.7\textwidth]{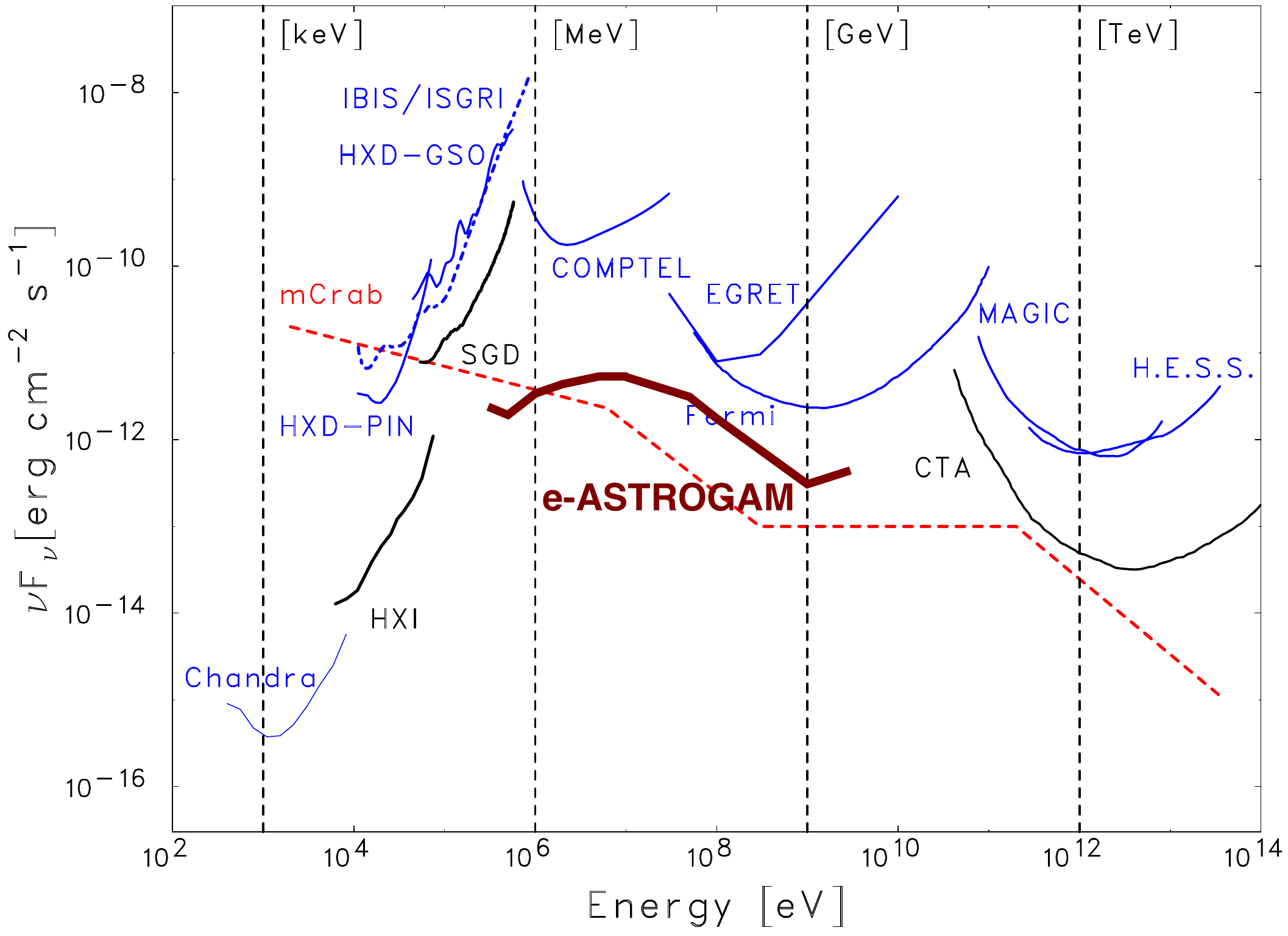}
\caption{Point source continuum sensitivity of different X and $\gamma$-ray instruments (adapted from Figure 1 in Ref.~\cite{tak13}). The curves for {\it Chandra}/ACIS-S, {\it Suzaku}/HXD (PIN, GSO), {\it INTEGRAL}/IBIS and {\it ASTRO-H} (HXI, SGD) are given for an observing time $T_{\rm obs}$ = 100 ks. The COMPTEL and EGRET sensitivities are given for the observing time accumulated during the whole duration of the {\it CGRO} mission ($T_{\rm obs}$ $\sim$ 9 years). The {\it Fermi}/LAT sensitivity is for a high Galactic latitude source and $T_{\rm obs}$ = 1 year \cite{atw09}. For MAGIC \cite{ale12}, H.E.S.S. \cite{aha06} and CTA \cite{act11} the sensitivities are given for $T_{\rm obs}$ = 50 hours. Sensitivities above 30 MeV are given at the 5-sigma confidence level, whereas those below 10 MeV (30 MeV for COMPTEL) are at 3-sigma. The e-ASTROGAM sensitivity is for an effective exposure of 1 year of a high Galactic latitude source.}
\label{fig:sensitivity}
\end{figure}

Improving the sensitivity in the medium-energy gamma-ray domain (1--100~MeV) by one to two orders of magnitude compared to previous missions is the main requirement for the proposed e-ASTROGAM mission. Such a performance will open an entirely new window for discoveries in the high-energy Universe.

Figure~\ref{fig:sensitivity} shows the computed e-ASTROGAM continuum sensitivity (assuming a spectral bin width $\Delta E = E$) for a 1-year effective exposure of a high Galactic latitude source, overlaid on a plot from Ref.~\cite{tak13} compiling the sensitivity of various X and gamma-ray instruments. Such an effective exposure will be reached for broad regions of the sky after 3 years of operation, given the very large field of view of the instrument (see Sect.~3.2). We see that e-ASTROGAM would provide an important leap in sensitivity over a wide energy band, from about 200 keV to 100 MeV. At higher energies, ASTROGAM would also provide a new vision of the gamma-ray sky thanks to its unprecedented angular resolution, which would reduce the source confusion that plagues the current {\it Fermi}-LAT and {\it AGILE} images near the Galactic plane (see, e.g., the 3FGL catalog\cite{3FGL}).

\subsection{Line sensitivity}

Table~\ref{tab:linesensitivity} shows the e-ASTROGAM 3$\sigma$ sensitivity for the detection of key gamma-ray lines from pointing observations, together with the sensitivity of the {\it INTEGRAL} Spectrometer (SPI). The latter was obtained from the {\it INTEGRAL} Observation Time Estimator (OTE) assuming 5$\times$5 dithering observations. The reported line widths are from SPI observations of the 511 and 847 keV lines (SN 2014J), and from theoretical predictions for the other lines. Noteworthy, the neutron capture line from accreting neutron stars can be significantly redshifted and broadened (FWHM between 10 and 100 keV) depending on the geometry of the mass accretion \cite{bil93}.

\begin{table}
\caption{e-ASTROGAM line sensitivity (3$\sigma$ in 10$^6$ s) compared to that of {\it INTEGRAL}/SPI\cite{roq03}.}
\label{tab:linesensitivity}       
\begin{tabular}{cccccc}
\hline\noalign{\smallskip}
E  & FWHM & Origin & SPI sensitivity & e-ASTROGAM sens. & Improvement \\
(keV) & (keV) & & (ph cm$^{-2}$ s$^{-1}$) & (ph cm$^{-2}$ s$^{-1}$) & factor\\
\noalign{\smallskip}\hline\noalign{\smallskip}
511 & 1.3 & Narrow line component of the & 5.2 $\times$ 10$^{-5}$ & 4.1 $\times$ 10$^{-6}$ & 13\\
  &  &  e+/e- annihilation radiation &  &  &  \\
  &  & from the Galactic center region &  &  &  \\
847 & 35 & $^{56}$Co line from thermonuclear& 2.3 $\times$ 10$^{-4}$ & 3.5 $\times$ 10$^{-6}$ & 66\\
 & & supernovae& & & \\
1157 & 15 & $^{44}$Ti line from core-collapse & 9.6 $\times$ 10$^{-5}$ & 3.6 $\times$ 10$^{-6}$ & 27\\
 & & supernova remnants & & & \\
1275 & 20 & $^{22}$Na line from classical novae& 1.1 $\times$ 10$^{-4}$ & 3.8 $\times$ 10$^{-6}$ & 29\\
 & &  of the ONe type & & & \\
2223 & 20 & Neutron capture line from& 1.1 $\times$ 10$^{-4}$ & 2.1 $\times$ 10$^{-6}$ & 52\\
 & & accreting neutron stars & & & \\
4438 & 100 & $^{12}$C line produced by low-energy & 1.1 $\times$ 10$^{-4}$ & 1.7 $\times$ 10$^{-6}$ & 65\\
 & & cosmic rays in the inner Galaxy& & & \\
\noalign{\smallskip}\hline
\end{tabular}
\end{table}

\begin{figure}
\centering
\includegraphics[width=0.55\textwidth]{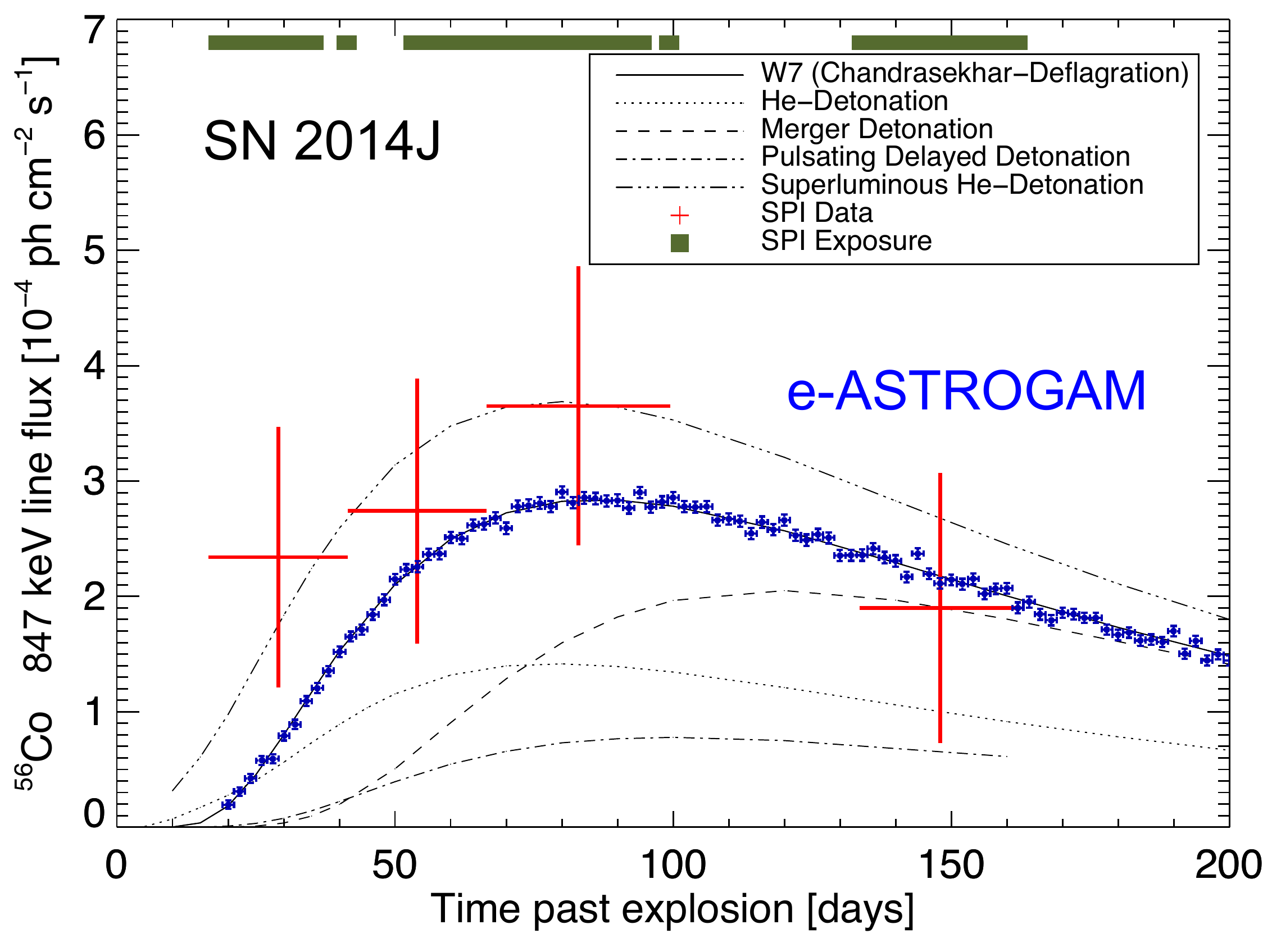}
\caption{Light curve of the 847 keV line from $^{56}$Co decay from the Type Ia supernova SN~2014J (adapted from Figure~4 in Ref.~\cite{die15}). The red data points are the observations of {\it INTEGRAL}/SPI and those in blue are from a simulation using the e-ASTROGAM response and assuming that the gamma-ray emission follows the W7 model \cite{nom84}. All the model light curves are from The \& Burrows (2014)\cite{the14}.}
\label{fig:sn2014j}
\end{figure}

We see that e-ASTROGAM will achieve a major gain in sensitivity compared to SPI for all gamma-ray lines, the most significant improvement being for the 847~keV line from Type Ia SNe. With the expected sensitivity of $3.5 \times 10^{-6}$ ph~cm$^{-2}$~s$^{-1}$ in 1~Ms of integration time (Table~\ref{tab:linesensitivity}), e-ASTROGAM should detect $10 \pm 3$ SNe~Ia in 3~years of nominal mission lifetime, up to a distance of $\sim$35~Mpc (for the brightest SNe). As illustrated in Figure~\ref{fig:sn2014j}, e-ASTROGAM will provide much better data than we have now with SPI for SN~2014J ($D=3.5\pm0.3$~Mpc) in case of similarly nearby events. These data will allow us to probe the explosion mechanism in detail, and compare with astrophysical models for each event to better understand the progenitor system(s) and the thermonuclear explosion process.

With the predicted line sensitivity (Table~\ref{tab:linesensitivity}), e-ASTROGAM will also (i) provide a much better map of the 511~keV radiation from positron annihilation in the inner Galaxy, (ii) uncover $\sim$10 young, $^{44}$Ti-rich SN remnants in the Galaxy and thus provide new insight on the explosion mechanism of core-collapse SNe (iii) detect for the first time the expected\cite{cla74} line from $^{22}$Na decay in novae hosted by ONe white dwarfs, (iv) provide a new constraint on the nuclear equation of state of neutron stars by detecting the predicted\cite{bil93} redshifted 2.2~MeV line from Scorpius X-1, and (iv) measure the energy density of low-energy cosmic rays in the inner Galaxy to better understand the role of these particles in the Galactic ecosystem.

\subsection{Polarization response}

Both Compton scattering and pair creation partially preserve the linear polarization information of incident photons. In a Compton telescope, the polarization signature is reflected in the probability distribution of the azimuthal scatter angle. In the pair domain, the polarization information is given by the distribution of azimuthal orientation of the electron-positron plane (see, e.g., Ref.\cite{ber50}). e-ASTROGAM will be able to perform unprecedented polarization measurements thanks to the fine 3D position resolution of both the Si Tracker and the Calorimeter, as well as the light mechanical structure of the Tracker, which is devoid of any heavy absorber in the detection volume.

\begin{figure}
\centering
\includegraphics[width=0.95\textwidth]{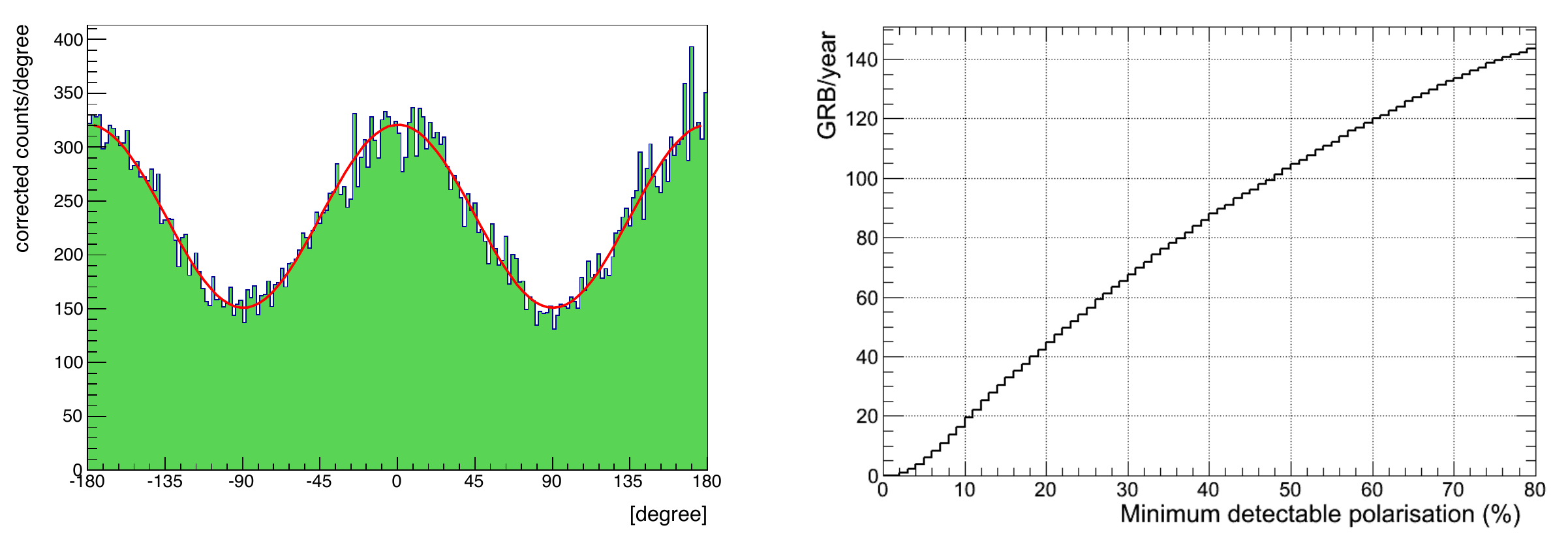}
\caption{{\it Left panel} -- e-ASTROGAM polarization response (polarigramme) in the 0.2 -- 2 MeV range for a 100\% polarized, 10 mCrab-like source observed on axis for 10$^6$ s. The corresponding modulation is $\mu_{100}$ = 0.36. {\it Right panel} -- Cumulative number of GRBs to be detected by e-ASTROGAM as a function of the minimum detectable polarization at the 99\% confidence level.}
\label{fig:polarization}
\end{figure}

The {\it left panel} of Figure~\ref{fig:polarization} shows an example of a polarigramme in the 0.2 -- 2 MeV range (i.e. in the Compton domain), simulated with MEGAlib. The calculations assume a 100\% polarized emission from a 10 mCrab-like source observed on axis. From the obtained modulation ($\mu_{100} = 0.36$), we find that at low energies (0.2 -- 2~MeV), e-ASTROGAM will be able to achieve a Minimum Detectable Polarization (MDP) at the 99\% confidence level as low as 0.7\% for a Crab-like source in 1~Ms (statistical uncertainties only). After one year of effective exposure of the Galactic center region, the achievable MDP$_{99}$ for a 10~mCrab source will be 10\%. With such a performance, e-ASTROGAM will be able to study the polarimetric properties of many pulsars, magnetars, and black hole systems in the Galaxy.

The {\it right panel} of Figure~\ref{fig:polarization} shows the number of GRBs detectable by e-ASTROGAM as a function of MDP$_{99}$ in the 150--300 keV band. The total number of GRBs detected by e-ASTROGAM will be $\sim$600 in 3 years of nominal mission lifetime. Here, the GRB emission spectrum has been approximated by a typical Band function\cite{ban93} with $\alpha=-1.1$, $\beta=-2.3$, and $E_{\rm peak}=0.3$~MeV, and the response of e-ASTROGAM to linearly polarized GRBs has been simulated at several off-axis angles in the range $[0^\circ;90^\circ]$. The number of GRBs with polarization measurable with e-ASTROGAM has then been estimated using the Fourth BATSE GRB Catalog\cite{pac99}. We see in Fig.~\ref{fig:polarization} that e-ASTROGAM should be able to detect a polarization fraction of 20\% in about 42 GRBs per year, and a polarization fraction of 10\% in $\sim$16 GRBs per year. This polarization information, combined with spectroscopy over a wide energy band, will provide unambiguous answers to fundamental questions on the sources of the GRB highly relativistic jets and the mechanisms of energy dissipation and high-energy photon emission in these extreme astrophysical phenomena.

\section{Summary}
\label{sec:summary}

e-ASTROGAM is a concept for a gamma-ray space observatory that has
the potential to revolutionize the astronomy of medium-energy
gamma rays by increasing the number of known sources in this field
by more than an order of magnitude and providing polarization
information for many of these sources. {\mt Thus, thousands of sources
are expected to be detected during the first 3 years of
operations.} Furthermore, the proposed wide-field gamma-ray
observatory 
{\mt will} play a major role in the development of time-domain
astronomy, and provide valuable information for the localization
and identification of gravitational wave sources.  
{\mt The instrument is} based on an innovative design, which
minimizes any passive material in the detector volume. The
instrument performance has been assessed through detailed
simulations using state-of-the-art simulation tools and the
results fully meet the scientific requirements of the proposed
mission. {\mt e-ASTROGAM will operate as an observatory open to
the international community: we envision an ESA-managed Guest
Observer Program for scientific investigations. The gamma-ray observatory
will be complementary to ground and space instruments, and
multi-frequency observation programs will be very important for
the success of the mission. In particular, e-ASTROGAM will be of
crucial importance for investigations jointly done with radio
(VLA, VLBI, ALMA, SKA), optical (JWST, E-ELT and other ground
telescopes), X-ray and TeV ground instrument (CTA, HAWK, LHAASO
and other ground-based detectors). Special emphasis will be given
to fast reaction to transients and rapid communication of alerts.
New astronomy windows of opportunity (sources of gravitational
waves, neutrinos, ultra high-energy cosmic rays) will be fully and
uniquely explored.}

\acknowledgments 
The research leading to these results has received funding from the European Union's Horizon 2020 Programme under the AHEAD project (grant agreement n. 654215).

\bibliography{spie2016_eastrogam_ref} 

\begin{thebibliography}{10}

\bibitem{3FGL}
{Acero}, F., {Ackermann}, M., {Ajello}, M., and {Fermi-LAT Collaboration},
  ``{Fermi Large Area Telescope Third Source Catalog},'' {\em \apjs}~{\bf 218},
   23 (June 2015).

\bibitem{bau13}
{Baumgartner}, W.~H., {Tueller}, J., {Markwardt}, C.~B., et~al., ``{The 70
  Month Swift-BAT All-sky Hard X-Ray Survey},'' {\em \apjs}~{\bf 207},  19
  (Aug. 2013).

\bibitem{sch00}
{Sch{\"o}nfelder}, V., {Bennett}, K., {Blom}, J.~J., et~al., ``{The first
  COMPTEL source catalogue},'' {\em \aaps}~{\bf 143},  145--179 (Apr. 2000).

\bibitem{tav16}
{Tavani}, M., {Tatischeff}, V., {von Ballmoos}, P., et~al., ``{The ASTROGAM
  gamma-ray space mission: A sensitive observatory for the MeV -- GeV
  domain},'' {\em Experimental Astronomy} ,  to be submitted (2016).

\bibitem{tav09}
{Tavani}, M., {Barbiellini}, G., {Argan}, A., et~al., ``{The AGILE Mission},''
  {\em \aap}~{\bf 502},  995--1013 (Aug. 2009).

\bibitem{atw09}
{Atwood}, W.~B., {Abdo}, A.~A., {Ackermann}, M., et~al., ``{The Large Area
  Telescope on the Fermi Gamma-Ray Space Telescope Mission},'' {\em \apj}~{\bf
  697},  1071--1102 (June 2009).

\bibitem{kan05}
{Kanbach}, G., {Andritschke}, R., {Zoglauer}, A., et~al., ``{Development and
  calibration of the tracking Compton/Pair telescope MEGA},'' {\em Nuclear
  Instruments and Methods in Physics Research A}~{\bf 541},  310--322 (Apr.
  2005).

\bibitem{bul07}
{Bulgarelli}, A., {Argan}, A., {Barbiellini}, G., et~al., ``{The AGILE silicon
  tracker: Pre-launch and in-flight configuration},'' {\em Nuclear Instruments
  and Methods in Physics Research A}~{\bf 614},  213--226 (Mar. 2010).

\bibitem{atw07}
{Atwood}, W.~B., {Bagagli}, R., {Baldini}, L., et~al., ``{Design and initial
  tests of the Tracker-converter of the Gamma-ray Large Area Space
  Telescope},'' {\em Astroparticle Physics}~{\bf 28},  422--434 (Dec. 2007).

\bibitem{per06}
{Perotti}, F., {Fiorini}, M., {Incorvaia}, S., {Mattaini}, E., and
  {Sant'Ambrogio}, E., ``{The AGILE anticoincidence detector},'' {\em Nuclear
  Instruments and Methods in Physics Research A}~{\bf 556},  228--236 (Jan.
  2006).

\bibitem{moi07}
{Moiseev}, A.~A., {Hartman}, R.~C., {Ormes}, J.~F., et~al., ``{The
  anti-coincidence detector for the GLAST large area telescope},'' {\em
  Astroparticle Physics}~{\bf 27},  339--358 (June 2007).

\bibitem{zog06}
{Zoglauer}, A., {Andritschke}, R., and {Schopper}, F., ``{MEGAlib The Medium
  Energy Gamma-ray Astronomy Library},'' {\em \nar}~{\bf 50},  629--632 (Oct.
  2006).

\bibitem{bul12}
{Bulgarelli}, A., {Fioretti}, V., {Malaguti}, P., {Trifoglio}, M., and
  {Gianotti}, F., ``{BoGEMMS: the Bologna Geant4 multi-mission simulator},'' in
  [{\em High Energy, Optical, and Infrared Detectors for Astronomy
  V}{\nolinebreak\hspace{0.1em}]},  {\em \procspie} {\bf 8453},  845335 (July
  2012).

\bibitem{cam14}
{Campana}, R., {Orlandini}, M., {Del Monte}, E., {Feroci}, M., and {Frontera},
  F., ``{The radiation environment in a low earth orbit:the case of
  BeppoSAX},'' {\em Experimental Astronomy}~{\bf 37},  599--613 (Nov. 2014).

\bibitem{tak13}
{Takahashi}, T., {Uchiyama}, Y., and {Stawarz}, {\L}., ``{Multiwavelength
  Astronomy and CTA: X-rays},'' {\em Astroparticle Physics}~{\bf 43},  142--154
  (Mar. 2013).

\bibitem{ale12}
{Aleksi{\'c}}, J., {Alvarez}, E.~A., {Antonelli}, L.~A., et~al., ``{Performance
  of the MAGIC stereo system obtained with Crab Nebula data},'' {\em
  Astroparticle Physics}~{\bf 35},  435--448 (Feb. 2012).

\bibitem{aha06}
{Aharonian}, F., {Akhperjanian}, A.~G., {Bazer-Bachi}, A.~R., et~al.,
  ``{Observations of the Crab nebula with HESS},'' {\em \aap}~{\bf 457},
  899--915 (Oct. 2006).

\bibitem{act11}
{Actis}, M., {Agnetta}, G., {Aharonian}, F., et~al., ``{Design concepts for the
  Cherenkov Telescope Array CTA: an advanced facility for ground-based
  high-energy gamma-ray astronomy},'' {\em Experimental Astronomy}~{\bf 32},
  193--316 (Dec. 2011).

\bibitem{bil93}
{Bildsten}, L., {Salpeter}, E.~E., and {Wasserman}, I., ``{Helium destruction
  and gamma-ray line emission in accreting neutron stars},'' {\em \apj}~{\bf
  408},  615--636 (May 1993).

\bibitem{roq03}
{Roques}, J.~P., {Schanne}, S., {von Kienlin}, A., et~al., ``{SPI/INTEGRAL
  in-flight performance},'' {\em \aap}~{\bf 411},  L91--L100 (Nov. 2003).

\bibitem{die15}
{Diehl}, R., {Siegert}, T., {Hillebrandt}, W., {Krause}, M., et~al., ``{SN2014J
  gamma rays from the $^{56}$Ni decay chain},'' {\em \aap}~{\bf 574},  A72
  (Feb. 2015).

\bibitem{nom84}
{Nomoto}, K., {Thielemann}, F.-K., and {Yokoi}, K., ``{Accreting white dwarf
  models of Type I supernovae. III - Carbon deflagration supernovae},'' {\em
  \apj}~{\bf 286},  644--658 (Nov. 1984).

\bibitem{the14}
{The}, L.-S. and {Burrows}, A., ``{Expectations for the Hard X-Ray Continuum
  and Gamma-Ray Line Fluxes from the Type Ia Supernova SN 2014J in M82},'' {\em
  \apj}~{\bf 786},  141 (May 2014).

\bibitem{cla74}
{Clayton}, D.~D. and {Hoyle}, F., ``{Gamma-Ray Lines from Novae},'' {\em
  \apjl}~{\bf 187},  L101 (Feb. 1974).

\bibitem{ber50}
{Berlin}, T.~H. and {Madansky}, L., ``{On the Detection of {$\gamma$}-Ray
  Polarization by Pair Production},'' {\em Physical Review}~{\bf 78},  623--623
  (June 1950).

\bibitem{ban93}
{Band}, D., {Matteson}, J., {Ford}, L., et~al., ``{BATSE observations of
  gamma-ray burst spectra. I - Spectral diversity},'' {\em \apj}~{\bf 413},
  281--292 (Aug. 1993).

\bibitem{pac99}
{Paciesas}, W.~S., {Meegan}, C.~A., {Pendleton}, G.~N., et~al., ``{The Fourth
  BATSE Gamma-Ray Burst Catalog (Revised)},'' {\em \apjs}~{\bf 122},  465--495
  (June 1999).

\end{thebibliography}
\bibliographystyle{spiebib} 

\end{document}